# Independent indistinguishable quantum light sources on a reconfigurable photonic integrated circuit

D.J.P. Ellis[1*], A.J. Bennett[1a], C. Dangel[1,b], J.P. Lee[1,c], J.P. Griffiths[1], T.A. Mitchell[2], T.-K. Paraiso[1], P. Spencer[2], D.A. Ritchie[2] and A.J. Shields[1].

[1] Toshiba Research Europe Limited, 208 Science Park, Milton Road, Cambridge, CB40GZ, UK.
[2] Cavendish Laboratory, Cambridge University, J. J. Thomson Avenue, Cambridge, CB3 0HE, UK
[a] now at School of Engineering, Cardiff University, Queen's Buildings, 14-17 The Parade, Cardiff CF24 3AA, UK.
[b] also at Physik Department, Technische Universitat Munchen, 85748 Garching, Germany.
[c] also at Department of Engineering, Cambridge University, 9 J.J. Thomson Avenue, Cambridge, CB3 0FA, UK.

**Abstract**

We report a compact, scalable, quantum photonic integrated circuit realised by combining multiple, independent InGaAs/GaAs quantum-light-emitting-diodes (QLEDs) with a silicon oxynitride waveguide circuit. Each waveguide joining the circuit can then be excited by a separate, independently electrically contacted QLED. We show that the emission from neighbouring QLEDs can be independently tuned to degeneracy using the Stark Effect and that the resulting photon streams are indistinguishable. This enables on-chip Hong-Ou-Mandel-type interference, as required for many photonic quantum information processing schemes.

Photonic Integrated Circuits (PICs) are rapidly becoming the default platform for photonic quantum applications due to their robustness, interferometric stability, reconfigurability and scalability[1,2,3]. Furthermore, they allow complex optical systems, which until recently would occupy a whole laboratory to be reduced to the size of a single chip.

PICs have been employed in many quantum applications including logic gates[2,4], higher order path entanglement[5], quantum walks[6,7], tests of Boson Sampling[8,9,10,11] on-chip quantum teleportation[12]. These PICs can support optical qubits encoded in path, time bin, polarisation or mode and can be interconverted[13,14]. All of these experiments have relied upon photons generated externally and delivered to the PICs through fibre.

One route to achieve on-chip photon generation is to use silicon waveguides to directly produce photons[15,16], but the Poissonian statistics of pair generation makes this inherently unscalable. An alternative scheme is to use quantum emitters with naturally sub-Poissonian photon statistics within a waveguide circuit made of the emitter's host material[17,18,19,20]. It is also possible to evanescently couple the emitter to a low-loss PIC[21,22,23] and these schemes have verified the emission of single photons. However, for realising large scale and compact circuits it will be necessary to integrate multiple sources of indistinguishable photons on the chip, as we demonstrate here.

The device we report utilises a "*plug, bond and play*" hybrid integration approach to passively align and join multiple, independent quantum light sources to a multi-channel PIC in an inherently scalable fashion. The Quantum Photonic Integrated Circuit, or QPIC, consists of three parts: (1) the array of Quantum Dot-containing light emitting diodes (QD LEDs), which are butt coupled to (2) a reconfigurable silicon oxynitride ($SiO_xN_y$) PIC on a silicon substrate and (3) an array of optical fibres for optical access to the device (see Figure **1**). We show the essential elements of a quantum-PIC including the generation of coherent, indistinguishable quantum light on-chip; the reconfigurability of the PIC; and the inclusion of an integrated excitation source.

The linear array of independently controllable QD LEDs can naturally emit single photons, photon pairs[24],[25] and using electric field tuning, transitions in separate emitters can be made degenerate.  The LEDs are designed to emit photons along the growth direction of the crystal and hence must be rotated through 90 degrees in order to butt-couple to the PIC. The challenge is then to obtain electrical contact from the LED array, as conventional bonding systems (such as wire bonders) are not designed to access out-of-plane contact pads. We achieve electrical connection using gold tracks on the surface, and a series of etched recesses at the left-hand end of the PIC. Raised metal contacts on the LEDs mate with these etched recesses to form a "plug and socket" arrangement, simultaneously aligning the LEDs and waveguides.  The whole device can then then be mounted on a simple "dip stick" and lowered directly into helium vapour, cooling the device to 8k. Under these conditions all of the LEDs and phase modulators can be operated simultaneously with minimal consumption of Helium.

The LED heterostructure was grown by molecular beam epitaxy and comprises a weak planar cavity with 4 (10) pairs of GaAs/AlGaAs distributed Bragg reflectors above (below) a $\lambda$-cavity spacer.  Self-assembled InAs quantum dots are placed at the centre of the cavity with an AlGaAs tunnel barrier on both sides to increase the range over which the quantum dot emission can be tuned. Mesas were wet etched and ohmic contacts produced by thermal evaporation and subsequent lift-off. The upper surface of the LED (p-doped) is partially covered with a gold contact, upon which a gold ball bump is placed.  This bump forms the male part of the "plug and socket" assembly scheme. To enable electrical contact to be made to the lower (n-doped) contact region, additional dummy mesas were added to the LED array and metal tracks defined running from the top of the dummy mesa down to the lower ohmic contact.

The photonic integrated circuit comprises a 1.6 µm thick core layer of n=1.55 $SiO_xN_y$, surrounded by n=1.51 $SiO_xN_y$ cladding layers such that the waveguide supports a single mode at our operation wavelength of around 900 nm.  All of the $SiO_xN_y$ layers are deposited by Plasma Enhanced Chemical Vapour Deposition (PECVD) onto a silicon substrate. Waveguides are defined lithographically and dry etched prior to the deposition of the overcladding layer. Further etching of the $SiO_xN_y$ and underlying silicon is then used to form thermal isolation trenches for the phase modulators and the "sockets" for the interconnection of the LEDs. Finally metal layers are deposited and patterned by lift-off to form the resistive heaters for the MZIs and the various contact tracks.

The fibre array is a commercially available component comprising an array of polarisation- maintaining (PM) optical fibres mounted in a v-groove array.  The fibres are used both to route in laser light (when off-chip excitation is employed) and to send quantum dot emission to detectors.

The fibre array and LED devices are bonded to the PIC using an index-matched UV-curable adhesive applied directly to the PIC's end facets. The LED array is aligned with the PIC by ensuring each gold ball bump is mated with the corresponding etched socket (as shown in Figure 1b).  This simultaneously aligns the electrical contacts and optical elements for each channel.  No active alignment is necessary. The presence of the plug and socket also prevents the bonding adhesive from electrically insulating the LED and PIC chips as any excess adhesive will fill from the bottom of the trench rather than flow over the raised contacts.  Finally, the remaining void between the gold bump and the etched trench is infilled with a silver-doped conductive epoxy allowing electrical contact from the surface of the PIC chip to the LEDs. The waveguides, LEDs and fibres are all arranged on a 250 µm pitch.

One section of the QPIC is as shown in Figure 2a, with two LEDs addressing a directional coupler. Figure 2b shows a typical IV curve. Photo-luminescence from each LED in the pair can be separately measured by biasing the unwanted LED at -5V to quench its emission (Figure 2c and d). Here the external 850 nm excitation laser is routed in through the waveguide circuit and pumps both LEDs simultaneously. Figure 2e shows a FEM calculation of the emitted field from an on-axis QD being guided into the waveguide. Only QDs aligned with the waveguide mode can contribute to the spectra. Also, the efficiency of light collection from the device is determined by the numerical aperture of the waveguide, giving a similar efficiency to what would be expected from a free-space lens of similar numerical aperture[26]. In the case shown in Figure 2e, we calculate a collection efficiency of 18%.

The second active element on the chip is the reconfigurable Mach-Zehnder interferometer (MZI), which contains a lithographically defined resistive heater (Figure 2f). Networks of phase shifters and MZIs can be used to realise any arbitrary circuit[3] and are used to prepare arbitrary qubit superposition states in rail-encoded QPIC schemes. In Figure 2g,h we drive one LED to emit electroluminescence into the MZI and use this to calibrate the device. Scanning the power applied to the phase shifter produces a clear switching between the output ports. At a wavelength of 890nm the couplers are closest to being balanced and we observe ideal switching operation (Figure 2i) with a visibility greater than 98% and a $2\pi$ phase shift requiring 520mW. Such high fidelity single qubit manipulation is an essential element of any versatile and reconfigurable QPIC.

The LEDs used here are designed to allow a giant Stark shift of >20nm[24] in reverse bias. In this regime no current flows, no electroluminescence is observed and optical excitation is required. All LEDs on the QPIC exhibit similar behaviour, as shown in the group of ten tuning spectra (Figure 3a).

We select two adjacent LEDs, 5 and 6, to demonstrate that photons from separated QDs can be rendered indistinguishable. Both devices were simultaneously optically excited through the waveguide circuit, and the QD photons collected through a single channel, as shown in Figure 3b. Figure 3c shows tuning of LED6 as the bias on LED5 is held constant at 0.92V and in Figure 3d the bias on LED5 is varied whilst LED6 is fixed at 0.89V. In both cases we are able to tune negatively charged transitions to degeneracy at 936.1nm when $V_{LED5}$ = 0.92 V and $V_{LED6}$ = 0.89 V. We then separately characterise the transitions, suppressing emission from the other LED by applying -5V. The transitions display a slightly different power dependence, so can be made to have the same intensity on the output channel when exciting with 126μW. At this point the linewidths of the transitions were measured and coherence times were determined to be 220 ps and 100 ps respectively (Figure 3f). Using a Hanbury Brown and Twiss setup and a pair of superconducting single photon detectors, the measured anti-bunching times of the two transitions in LED5 (LED6) were 913ps (927ps) and $g^{(2)}(0)$ was 0.230 (0.269) (Figure 3g,h). The measured non-zero $g^{(2)}(0)$ may be attributed to background emission in the LED arising from other QDs in the device.

We then perform a Hong-Ou-Mandel experiment to study on-chip two photon interference. With both devices tuned to degeneracy, if indistinguishable, the photons from the two QDs will undergo two-photon interference at the directional coupler and both leave from the same output port. Therefore, measuring coincidences between the two detectors in Figure 4a shows a "peak" in the autocorrelation function. This is precisely what we measure (Figure 4b).

Using the independently measured characteristics for the two LEDs we calculate the expected autocorrelation function, shown as red lines in Figure 4. The model agrees well with the experimental data. For comparison, the blue curve shows the results of a similar calculation in which no interference takes place. When we introduced a small energy difference between the transitions, the visibility of interference falls sharply as predicted (green curve in Figure 4c,d). At zero detuning, and with infinitely fast detectors, we predict a maximum two photon interference visibility of 80% limited by the non-zero $g^{(2)}(0)$ of the sources. Taking into account the measured detector response, we predict a reduced visibility of 54%.

Finally, we demonstrate a route to make our QPIC's source completely self-contained and scalable. In Figure 5 we show how the LED array itself can also be used to simultaneously excite and tune the energy of dots, without an external laser[27]. The lower (Master, *M*) LED in Figure 5a is driven at 10mA generating ~880nm wetting layer emission that was guided laterally by the cavity spacer. This excited QDs in the adjacent devices (Slave LEDs *S1*, *S2*), which can be tuned. Figure 5b shows experimental data recorded with this on-chip pumping scheme. Light collected from the upper LED (*S2*, held at fixed bias) occurs at constant wavelength, but light from the lower LED (*S1*, biased with $V_{DC}$) displays the giant stark shift. With the current LED array design, some of the light from the pump LED is also collected by the waveguide. The resulting background can be eliminated through future modifications to the device design. This all electrical, tunable excitation scheme therefore offers a route to a truly independent and scalable all electrical QPIC.

The integration of a tuneable single photon LED array with a reconfigurable photonic integrated circuit is a step forward in the direction of chip-scale quantum information processing with on-demand, high fidelity quantum sources. Future implementations of the current prototypes on low QD density wafers will eliminate the finite background in $g^{(2)}(\tau)$ measurements. We evaluate that a 10dB reduction of the background would lead to a 2-photon interference visibility of 98**%**, sufficient to allow high fidelity on-chip quantum logic gates such as the C-NOT gate to be realised**.** Further optimisation of the waveguide profile at the LED/PIC interface, and in the longer term, deterministic QD positioning[28] will provide additional filtering, reducing the contribution of off-axis QDs and improve the collection efficiency. We also expect greater indistinguishability from resonance fluorescence with the exciting light guided through the LED cavity[29]. Together, these improvements will allow this scalable platform to be deployed in a wide range of new applications.

In conclusion, we have demonstrated a compact, packaged device that integrates a reconfigurable loss-loss photonic circuit with an array of independently controlled single quantum dot light sources. Choosing a QLED design that allows substantial changes to the energy of transitions in the single quantum dots, we are able to show that photons from separate quantum dots can be made indistinguishable. This is a key requirement of linear optical quantum computing schemes.

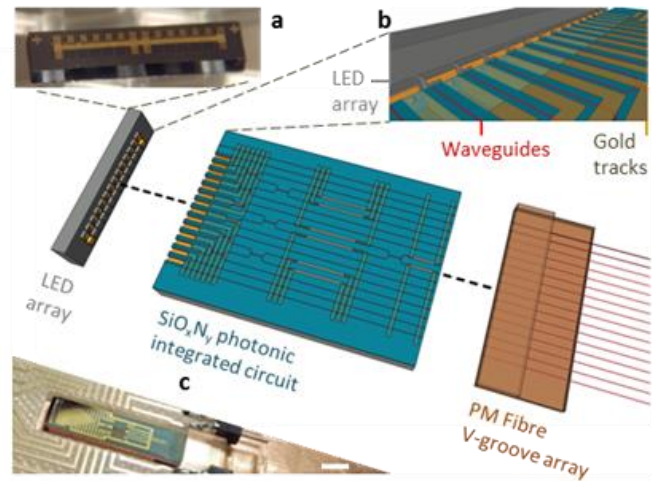

Figure 1. Assembly of the Quantum Photonic Integrated Circuit. The main figure shows an exploded diagram of the LED array, $SiO_xN_y$ waveguide circuit and PM fibre array. **a.** Optical photograph of the LED array (scale bar is 2mm). **b.** Schematic of the bonding interface between the LED array and waveguide circuit. **c.** Optical photograph of the assembled device (scale bar is 5mm).

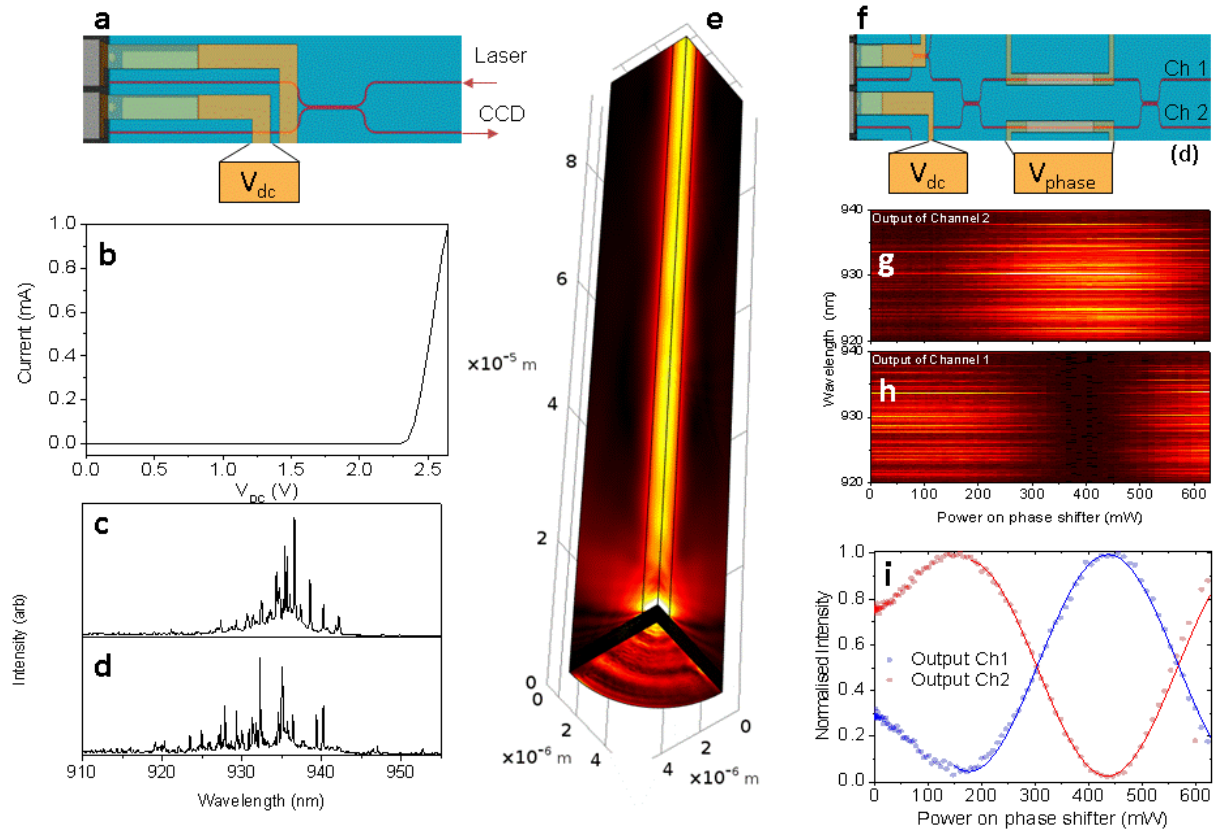

Figure 2. Activating and redirecting light on the Quantum Photonic Integrated Circuit. **a.** Schematic of the operation of the device under external optical excitation. **b.** Typical IV curve. **c.** Photoluminescence spectra recorded from two LEDs at 1.0V. **c,d.** Schematic showing operation of the on-chip phase shifters. **e.** FEM simulation showing coupling between an on-axis quantum dot in a low-quality planar cavity and a butt-coupled straight waveguide. **f.** Electroluminescence from one LED measured through channels 1 and 2 respectively as the power applied to the phase shifter is varied. A clear switching behaviour is observed. **g.** Single wavelength (890nm) data from e. replotted in more detail. We observe ~ideal MZI behaviour.

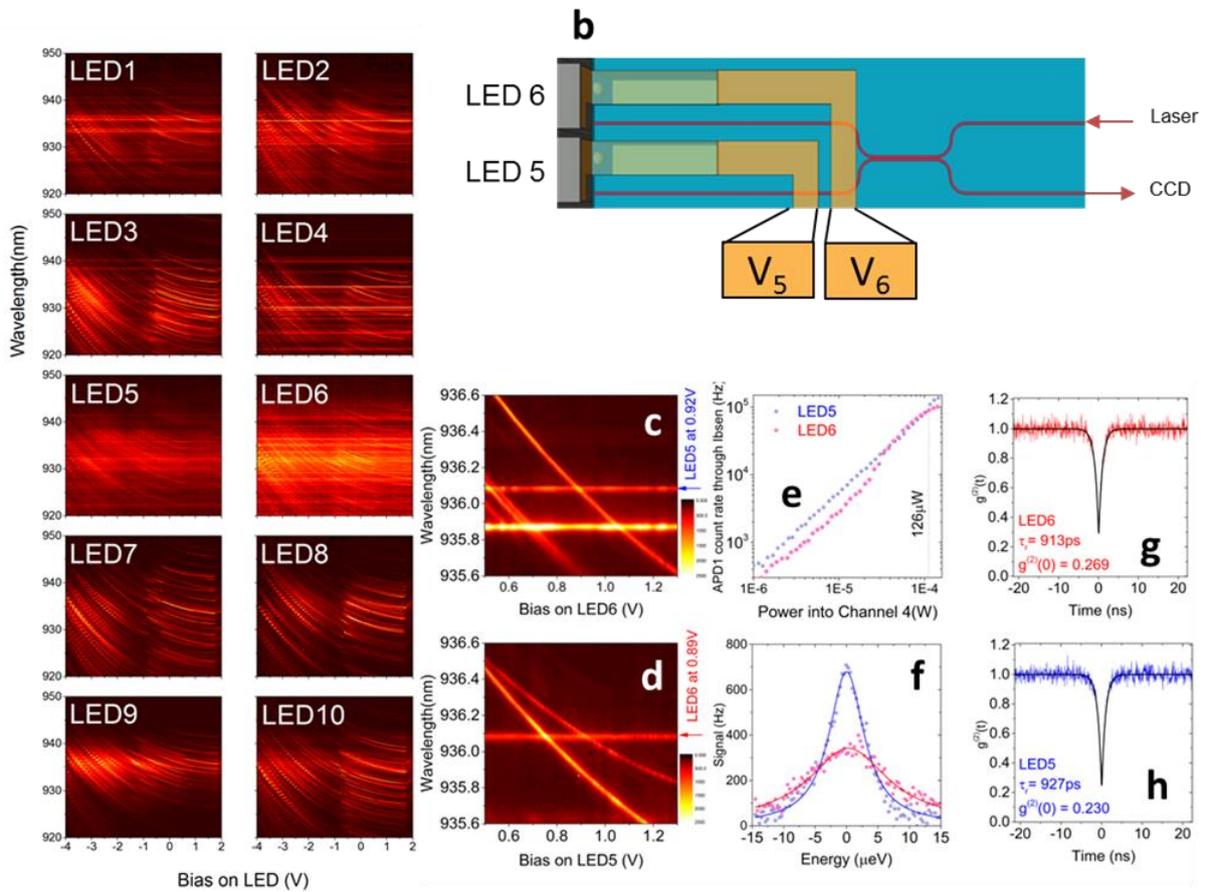

Figure 3. Making the transitions in separated emitters degenerate. **a.** Photoluminescence vs bias scans of all 10 LEDs on the device. Also show similar tuning behaviour. **b**. Schematic of the device with LEDs 5 and 6 studied in more detail. **c.** PL tuning curve. LED 5 held at 0.92 V. Bias applied to LED 6 scanned. **d.** PL tuning curve. As for c. except LED 5 is now scanned and LED 6 is held constant. **e.** Photoluminescence power dependence data for the transitions at degeneracy. **f.** Etalon linewidth measurements at degeneracy. **g, h.** Autocorrelation histograms for LEDs 5, 6 respectively.

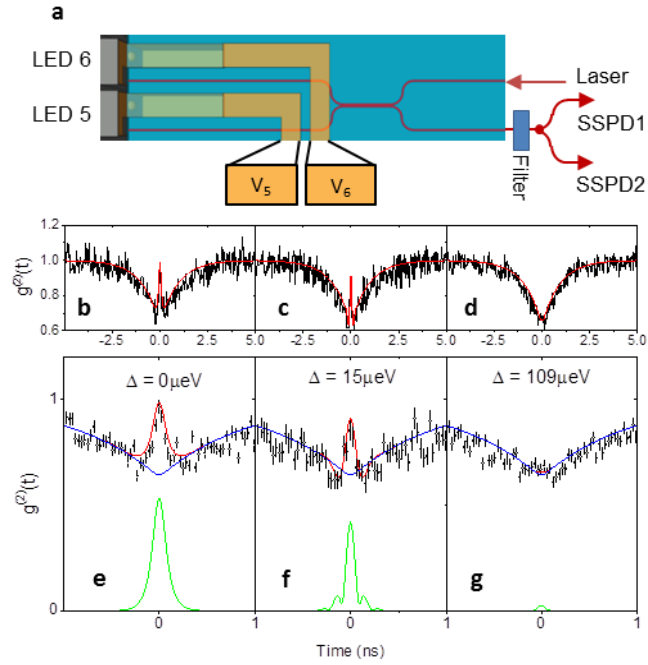

Figure 4. On-chip two photon interference of light from separate LEDs. **a**. The experimental setup for testing the indistinguishability of light from LED5 and LED6. **b-d.** Experimental Hong-Ou-Mandel histograms for detunings of 0, 15 and 109 µeV between the two quantum dot transitions, which are enlarged in **e-g**. The red (blue) line is a calculation based on experimental parameters when maximum (minimum) indistinguishability is present. The green line plots the predicted visibility.

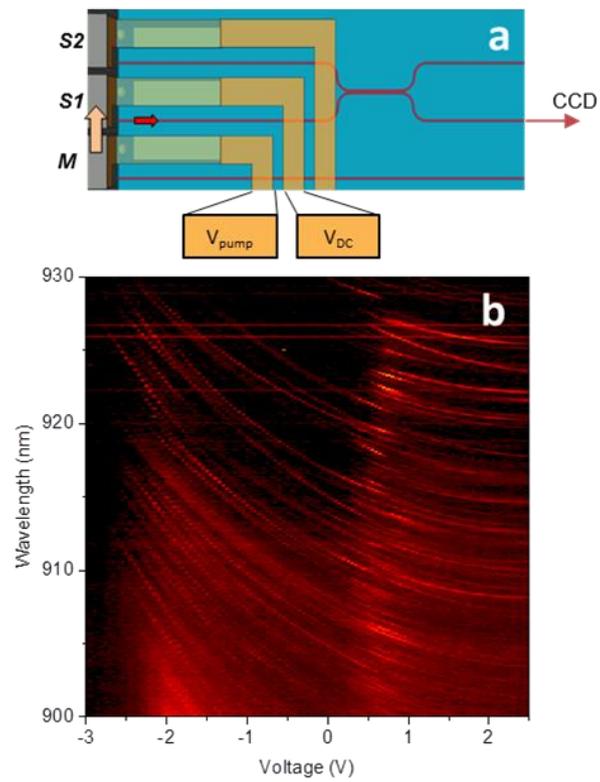

Figure 5. Integrating the excitation source onto the QPIC. **a.** Schematic of the section of the device used to demonstrate simultaneous on-chip pumping and tuning of a quantum emitter. A bias of $V_{pump}$ is applied to the Master LED (*M*), and strong emission at 880nm is guided to two adjacent Slave LEDs (*S1, S2*). One of these is tuned by bias $V_{DC}$ to produce the plot in **b.**

**Acknowledgements**

We thank E. Murray and T. Meany for assistance at an early stage of this project, R. M. Stevenson for assistance with the superconducting detectors and Kevin and Trevor for hardware support. We acknowledge funding from the EPSRC for MBE the system used for the growth of the LED. C.D. acknowledges support from the Marie Curie Actions within the Seventh Framework Programme for Research of the European Commission, under Network PICQUE (Grant No. 608062). P.S. acknowledges support from the EPSRC Quantum Communications Hub (Grant No EP/M013472/1).